\pdfoutput=1
\documentclass[pre,twocolumn,showpacs,superscriptaddress,aps]{revtex4}

\usepackage{amsfonts}
\usepackage{amsmath}
\usepackage{amssymb}
\usepackage{graphicx}
\usepackage{dcolumn}
\usepackage{times}

\DeclareMathOperator{\sech}{sech}

\begin{document}

\title{Quasi-discrete microwave solitons in a split ring resonator-based left-handed coplanar waveguide}
\author{G.P. Veldes}
\affiliation{Department of Physics, University of Athens, Panepistimiopolis,
Zografos, Athens 15784, Greece}
\author{J. Cuevas}
\affiliation{Grupo de F\'{\i}sica No Lineal, Departamento de F\'{\i}sica Aplicada I,
Escuela Polit\'ecnica Superior, C/ Virgen de \'{A}frica, 7, 41011
Sevilla, Spain}
\author{P.G.\ Kevrekidis}
\affiliation{Department of Mathematics and Statistics, University of Massachusetts,
Amherst MA 01003-4515, USA}
\author{D.J.\ Frantzeskakis}
\affiliation{Department of Physics, University of Athens, Panepistimiopolis,
Zografos, Athens 15784, Greece}

\begin{abstract}
We study the propagation of quasi-discrete microwave solitons in a nonlinear left-handed coplanar waveguide coupled with split ring resonators. By considering the relevant transmission line analogue, we derive a nonlinear lattice model which is studied analytically by means of a quasi-discrete approximation. We derive a nonlinear Schr{\"o}dinger equation, and find that the system supports bright envelope soliton solutions in a relatively wide subinterval of the left-handed frequency band. We perform systematic numerical simulations, in the framework of the nonlinear lattice model, to study the propagation properties of the quasi-discrete microwave solitons. Our numerical findings are in  good agreement with the analytical predictions, and suggest that the predicted structures are quite robust and may be observed in experiments.

\end{abstract}


\maketitle

\section{Introduction}

Artificially engineered {\it metamaterials} have unique electromagnetic (EM) properties, which are of great interest both from the basic physics viewpoint and for numerous applications \cite{review1,review2,review3}. In such media, the  effective permittivity $\epsilon$ and permeability $\mu$ is such that, in certain frequency bands, the metamaterial displays either a right-handed (RH) behavior ($\epsilon>0$, $\mu>0$) or a left-handed (LH) behavior ($\epsilon<0$, $\mu<0$); in other words, energy and wavefronts may travel in the same or opposite directions in the RH or the LH frequency region, respectively. Metamaterials with a LH behavior, namely LH-metamaterials (LHM), exhibit negative refraction at microwave \cite{exp1,exp2} or optical frequencies \cite{shalaev}.

Apart from the effective medium description, there has also been a large interest in studying equivalent  transmission line (TL) models for LHMs in the microwave frequency region. In such a case, the effective permittivity $\epsilon $ and permeability $\mu$ are directly connected to the serial and shunt impedance of the TL model; this way, so-called, {\it composite right-left handed} (CRLH) TLs \cite{Caloz1} may exhibit either a RH or a LH behavior depending on the frequency band. In practice, CRLH-TLs may be implemented with the coplanar waveguide (CPW)~\cite{Eleftheriades} or microstrip technology~\cite{Caloz2}. Importantly, CRLH-TLs have led to many microwave applications and devices, including dual-band branch-line couplers, asymmetric backward-wave directional couplers, resonators, antennas, and so on \cite{review1,review2,review3,Caloz1}.

On the other hand, {\it nonlinear} metamaterials, namely structures in which $\epsilon$ and $\mu$ (or the serial and shunt impedance in the respective TL models) depend on the EM field intensities (or voltages and currents in the TL models), have also been a subject of interest. Such structures may be implemented by embedding an array of wires and split ring resonators (SRRs) into a nonlinear dielectric \cite{zharov,agranovich}, or by inserting diodes into resonant conductive elements (such as the SRRs) \cite{lapine,lapine2,soukoulis,ysk}. Nonlinear metamaterials may prove useful in various applications, including ``switching'' the material properties from left- to right-handed and back, tunable structures with intensity-controlled transmission, negative refraction photonic crystals, etc. Furthermore, fundamental effects, such as harmonic generation, nonlinearity-induced localization of EM waves and soliton formation, are possible. More specifically, relevant nonlinear phenomena \cite{feng,chowd} and soliton formation have already been predicted to occur in nonlinear metamaterials, using either the effective medium description (see, e.g., Refs.~\cite{laz-ts,kourakis,shukla,scalora,wen-pre,wen-pra,tsitsas1,tsitsas2}) or the TL description (see, e.g., Refs.~\cite{koz1,nar,cam}). From the viewpoint of nonlinear TL experiments, pulse propagation \cite{koz2} and envelope soliton formation \cite{koz3} were recently observed (see also the review of Ref.~\cite{revtl} and the relevant work of Ref.~\cite{lars}); moreover, analytical approximations, based on a continuum nonlinear Schr\"{o}dinger (NLS) equation, allowed the description of bright \cite{ogas,lars} or dark \cite{wang,lars} envelope solitons observed in the experiments.

In this work, we study {\it quasi-discrete} microwave solitons that may be formed in a {\it planar} LHM. Such structures have the advantage of being easily fabricated (by means of standard mask/photoetching techniques), they are compact, and are compatible with monolithic microwave integrated circuits. At this point we should mention that, generally, metamaterial TLs, operating at microwave frequencies, are artificial lines consisting of a host line loaded with reactive elements. Such TLs can be implemented by means of two main approaches: (i) the so-called ``CL-loaded approach'', where RH-TLs are loaded with series capacitances and shunt inductances \cite{Caloz1,Eleftheriades,Caloz2} (see also Ref.~\cite{lars} for results in a nonlinear case), and (ii) the so-called ``resonant-type approach'', where the TLs are loaded with sub-wavelength resonators, such as SRRs \cite{Martin,mart2}. TLs of the latter type exhibit controllable electrical characteristics, beyond what can be achieved in conventional TLs, implemented, e.g., in printed circuit boards (PCBs): the size of such a TL is determined by the size of the resonators and, thus, they can easily be miniaturized. Note that the fabrication of a pertinent prototype device, based on a CPW with an array of SRRs being etched at the bottom of the substrate, was first introduced in Ref.~\cite{Martin}, while the corresponding TL model was presented in Ref.~\cite{Aznar}.

Here, we consider a nonlinear counterpart of the TL model of Ref.~\cite{Aznar}, with the shunt capacitors of the model being nonlinear; such a nonlinear TL model may be implemented by incorporating a nonlinear dielectric thin film in the structure, whose dielectric constant may be controlled by a proper bias voltage (see, e.g., Refs.~\cite{findi1,findi2,cai}). In our analysis, starting from the discrete lump element model of the CRLH-TL under consideration, we derive a nonlinear lattice equation, which is then treated in the framework of the {\it quasi-discrete} (alias {\it quasi-continuum}) approximation (see, e.g., Ref.~\cite{Marquie} and Ref.~\cite{Remoissenet} for a review): this way, seeking for envelope soliton solutions of the nonlinear lattice model, characterized by a {\it discrete carrier} and a {\it continuum envelope}, we employ a multi-scale perturbation method to derive an effective NLS equation. The coefficients of this equation, which determine the type (bright or dark) of the envelope soliton, are found and it is shown that bright NLS solitons are supported by the SRR-CPW structure in a relatively wide range of frequencies inside the LH frequency band. Our analytical predictions are corroborated by numerical simulations, which reveal (apart
from the basic properties) the robustness of the predicted quasi-discrete microwave solitons.

\begin{figure}[tbp]
\centering
\includegraphics[width=8cm]{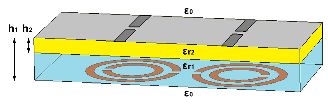}
\includegraphics[width=8cm]{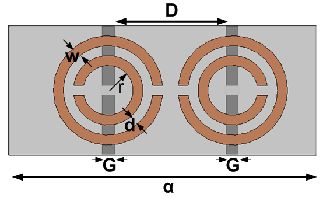}
\caption{(Color online) Top panel: Configuration of the SRR-CPW structure: the dielectric layer of relative dielectric constant $\epsilon_{r1}$ and width $h_1-h_2$ is linear, while the one of relative dielectric constant $\epsilon_{r2}$ and width $h_2$ is nonlinear. Bottom panel: Layout of the bottom plate of the configuration, where the SRRs are placed. Shown are all relevant dimensions appearing in the equations of Appendix A.}
\label{cpwsrr}
\end{figure}

The paper is organized as follows: in Sec.~II we present the SRR-CPW nonlinear structure and derive a nonlinear lattice model describing the evolution of the voltage across this structure; then we employ the quasi-discrete approximation and derive the NLS equation describing envelope solitons that can be supported in this setting. In Sec.~III we present numerical simulations in the framework of the nonlinear lattice model, reveal the propagation properties of the predicted quasi-discrete solitons, and compare the numerical results with the analytical predictions. We also propose changes in the configuration to increase the domains of existence of solitons. Finally, in Sec.~IV we discuss our conclusions.
Our presentation is complemented by three appendices: the first details
the parameters of the SRR-CPW model, the second analyzes the method
of multiple scales used to obtain the relevant NLS equation, while the
third touches upon the continuum limit of the quasi-discrete approximation
developed herein.

\section{The model and its analytical consideration}

\subsection{The nonlinear lattice model}

\begin{figure}[tbp]
\centering
\includegraphics[width=8cm]{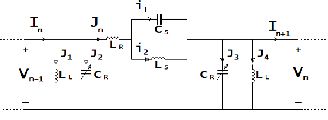}
\caption{The unit cell circuit of the SRR-CPW model.}
\label{fig:circuit model}
\end{figure}

We start by considering a nonlinear version of the SRR-CPW model introduced in Refs.~\cite{Martin,Aznar}, as shown in the top panel of Fig.~\ref{fig:circuit model}: here, the structure incorporates a nonlinear dielectric film, of relative dielectric constant $\epsilon_{r2}$ and width $h_2$. The rest of the SRR-CPW configuration is identical to the one of Refs.~\cite{Martin,Aznar}: one may observe the SRRs (of external radius $R$) at the bottom plate of the structure, which are aligned with the slots (of width $G$ and separation distance $D$) at the top plate of the structure. Notice that below we will focus on the case where the nonlinear dielectric film is introduced as shown in the top panel of Fig.~\ref{fig:circuit model}; according to this consideration, nonlinearity is only introduced in certain elements (i.e., shunt capacitors) in the equivalent discrete unit-cell model of the system. Nevertheless, below we will first consider the case where the serial capacitors, associated with the SRRs, are also nonlinear; in practice, this can be done, e.g., by inserting diodes in the SRR slots \cite{lapine,lapine2,soukoulis,ysk}.

The discrete element model (unit cell) of the considered SRR-CPW structure is shown in Fig.~\ref{fig:circuit model}. Here, $L_{R}$ and $C_{R}$ denote the equivalent per section inductance and capacitance of the line, respectively, $L_{s}$ and $C_{s}$ are the equivalent inductance and capacitance of the SRR, which is coupled with the transmission line, while the inductance $L_{L}$ is the equivalent inductance of the shunt strips. The above elements are directly connected with the physical parameters of the SRR-CPW structure (see, e.g., Ref.~\cite{chenchu} and details in Appendix A).

Let us now consider Kirchhoff's voltage and current laws for the SRR (see the $L_{s}$$C_{s}$ combination in Fig.~\ref{fig:circuit model}) equivalent circuit, namely,
\begin{eqnarray}
L_{s}\frac{di_{2}}{dt}&=&V_{n-1}-V_{n}-L_{R}\frac{dJ_{n}}{dt},
\label{eq:Kirch01}
\\
i_{1}&=&C_{s}\frac{d}{dt}\left(V_{n-1}-V_{n}-L_{R}\frac{dJ_{n}}{dt} \right).
\label{eq:Kirch02}
\end{eqnarray}
The above equations, together with the auxiliary Kirchhoff's current law $i_{1}+i_{2}=J_{n}$, lead to the following system for the unknown voltages $V_{n}$ and $U_n$:
\begin{eqnarray}
&&L_{R}\frac{d^{2}}{dt^{2}}[C_{s}(U_{n})U_{n}-C_{s}(U_{n+1})U_{n+1}] \nonumber \\
&&+\left(1+\frac{L_{R}}{{L_s}}\right)( U_{n}-U_{n+1})
+(V_{n-1}-2V_{n}+V_{n+1})=0, \nonumber \\
\label{eq:sys01}
\\
&&L_{R}\frac{d^{2}}{dt^{2}}[C(V_{n})V_{n}]-(V_{n-1}-2V_{n}+V_{n+1})\nonumber \\
&&+\frac{L_{R}}{L}V_{n}-( U_{n}-U_{n+1})=0,
%\nonumber \\
\label{eq:sys02}
\end{eqnarray}
where $C = 2 C_{R}$ and $L ={L_{L}}/{2}$. At this point, in order to further simplify Eqs.~(\ref{eq:sys01})-(\ref{eq:sys02}), we need to make some considerations concerning the nonlinear (voltage-dependent) capacitance $C=C(V_n)$ and $C_s(U_n)$. As discussed above, $C$ is effectively filled by a nonlinear dielectric, whose dielectric constant may be controlled by a proper bias voltage, say $V_{B}$ (see, e.g., Refs.~\cite{findi1,findi2,cai} and discussion below). On the other hand, the effective nonlinearity of $C_s$ is due to the presence of a diode, biased at a constant voltage, say $U_B$. Assuming that the voltages $V_{n}$ and $U_n$, respectively applied in the effective capacitors $C$ and $C_s$, do not change significantly from their relevant bias voltages $V_0$ and $U_0$, we can Taylor expand $C(V_n)$ and $C_s(U_n)$ taking into regard only the lowest order terms, namely,
\begin{eqnarray}
&&C(V_{B}+V_{n}) \approx C_{0}+C_{0}^{'}V_n,
%+\frac{1}{2}C_{0}^{''}V_n^{2}
\label{eq:Taylor1}
\\
&&C_s(U_{B}+U_{n}) \approx C_{s0}+C_{s0}^{'}U_n,
%+\frac{1}{2}C_{s0}^{''}U_n^{2},
\label{eq:Taylor2}
\end{eqnarray}
where $C_{0} \equiv C(V_{B})$ and $C_{s0} \equiv C_s(U_{B})$ are constant capacitances corresponding to $V_{B}$ and $U_B$, respectively, and similarly $C_{0}^{'} \equiv (dC/dV)\mid_{V_{B}}$,
%$C_{0}^{''} \equiv (d^2C/dV^2)\mid_{V_{0}}$,
$C_{s0}^{'} \equiv (dC_s/dU)\mid_{U_{B}}$,
%$C_{s0}^{''} \equiv (d^2C_s/dU^2)\mid_{U_{0}}$
take constant values depending on the particular forms of $C(V)$ and $C_s(U)$ (see also below).
Next, substituting Eqs.~(\ref{eq:Taylor1})-(\ref{eq:Taylor2}) into Eqs.~(\ref{eq:sys01})-(\ref{eq:sys02}), and introducing the scale transformations $t \rightarrow \omega_{sh} t$ [where $\omega_{sh}^{2}= (LC_{0})^{-1}$] and
$\{V_n, U_n \} \rightarrow [C_{0}^{'}(2C_{0})^{-1}]\{V_n, U_n \}$, we cast the system of Eqs.~(\ref{eq:sys01})-(\ref{eq:sys02}) into the following form:
\begin{eqnarray}
&&\frac{d^{2}V_{n}}{dt^{2}}-\beta^2(V_{n+1}+V_{n-1}-2V_{n})+V_{n} +\frac{d^{2}V_{n}^{2}}{dt^{2}}
\nonumber\\
%+\delta\frac{d^{2}V_{n}^{3}}{dt^{2}}
&&=\beta^2( U_{n}-U_{n+1}),
\label{eq:mod01}
\end{eqnarray}
and
\begin{eqnarray}
&&\frac{d^{2}}{dt^{2}}( U_{n}-U_{n+1})+\gamma^2( U_{n}-U_{n+1})+\sigma\frac{d^{2}}{dt^{2}} (U^{2}_{n}-U^{2}_{n+1})
\nonumber\\
&&+(\gamma^2-\mu^2) (V_{n+1}+V_{n-1}-2V_{n}) =0,
%\nonumber\\
%&+&\sigma\frac{d^{2}}{dt^{2}} (U^{2}_{n}-U^{2}_{n+1})=0,
%\nonumber\\
%&+&\frac{2}{3}\sigma \nu \frac{d^{2}}{dt^{2}}( U^{3}_{n}-U^{3}_{n+1})=0,
\label{eq:mod02}
\end{eqnarray}
where the constant parameters are given by:
\begin{eqnarray}
\sigma=\frac{C_{0}C_{s0}^{'}}{C_{0}^{'}C_{s0}}, \,\,\,\
%\delta=\frac{2}{3}\frac{ C_{0}C_{0}^{''}}{(C_{0}^{'})^{2}}, \,\,\,\
%\nu = \frac{C_{0}C_{s0}^{''}}{C_{0}^{'}C_{s0}^{'}}
%\nonumber \\
%&&
\gamma=\frac{f_{se}}{f_{sh}}, \,\,\,\
\beta=\frac{f_{RH}}{f_{sh}}, \,\,\,\
\mu=\frac{f_{s}}{f_{sh}}.
\label{eq:param1}
\end{eqnarray}
In the above expressions, $f_{se}$ and $f_{sh}$ denote series and shunt frequencies, while $f_s$ and $f_{RH}$ denote characteristic frequencies related to the SRR and the RH part of the circuit, respectively; the above frequencies read:
\begin{eqnarray}
f_{se}&=&\frac{1}{2 \pi }\sqrt{\frac{1}{L_{s}C_{s}}+\frac{1}{L_{R}C_{s}}},\qquad
f_{sh}=\frac{1}{2 \pi \sqrt {LC_{0}}},
\nonumber \\
f_{s}&=&\frac{1}{2\pi \sqrt {L_{s}C_{s}}}, \qquad f_{RH}=\frac{1}{2 \pi \sqrt {L_{R}C_{0}}}.
\label{eq:freq}
\end{eqnarray}

The system of Eqs.~(\ref{eq:mod01})-(\ref{eq:mod02}) is a model of two coupled nonlinear lattice equations, for the unknown voltages $V_n$ and $U_n$, describing the dynamics of the system. The analysis of the above system by means of analytical (and/or even numerical) techniques is a far more involved task and will
be deferred to a future study. In fact, we choose to consider a variant of the above model, which is analytically (and numerically) more tractable, corresponding to the case where only the capacitance $C$ (associated with the CPW structure) is nonlinear, while the capacitance $C_s$ is linear (i.e., nonlinearity in the considered SRR-CPW structure is only introduced by the insertion of the nonlinear dielectric film, as shown in Fig.~\ref{cpwsrr}). In such a case, the parameter $\sigma$ (governing the SRR nonlinearity) becomes $\sigma=0$, and the system of Eqs.~(\ref{eq:mod01})-(\ref{eq:mod02}) can be decoupled, leading to a single nonlinear lattice equation for the function $V_n$, namely:
%
%\begin{eqnarray}
%&& \frac{d^{4}V_{n}}{dt^{4}}+(1+\gamma^{2})\frac{d^{2}V_{n}}{dt^{2}}
%-\beta^2\frac{d^{2}}{dt^{2}}(V_{n+1}+V_{n-1}-2V_{n})\nonumber\\
%&-&\mu^2\beta^2(V_{n+1}+V_{n-1}-2V_{n})+\gamma^{2}V_{n}+\gamma^{2}
%\frac{d^{2}V_{n}^{2}}{dt^{2}}\nonumber\\
%&+&\frac{d^{4}V_{n}^{2}}{dt^{4}}+\delta\gamma^{2}
%\frac{d^{2}V_{n}^{3}}{dt^{2}} +\delta\frac{d^{4}V_{n}^{3}}{dt^{4}}=0.
%\label{eq:genmodel}
%\end{eqnarray}
%
%Finally, when $\delta=0$
%
\begin{eqnarray}
&& \frac{d^{4}V_{n}}{dt^{4}}+(1+\gamma^{2})\frac{d^{2}V_{n}}{dt^{2}}
-\beta^2\frac{d^{2}}{dt^{2}}(V_{n+1}+V_{n-1}-2V_{n})\nonumber\\
&-&\mu^2\beta^2(V_{n+1}+V_{n-1}-2V_{n})+\gamma^{2}V_{n}
\nonumber\\
&+&\gamma^{2}
\frac{d^{2}V_{n}^{2}}{dt^{2}} +\frac{d^{4}V_{n}^{2}}{dt^{4}}=0.
\label{eq:model}
\end{eqnarray}

It is now useful, for the purposes of our analytical and numerical considerations, to adopt experimentally relevant parameter values. First, following Ref.~\cite{Aznar}, the parameters related to the CPW structure are chosen as follows.
The waveguide has a width $a=23.7$~mm and a thickness $h_1=1.27$~mm, the central strip width is $D=7$~mm, the width of the slots is $G=1.35$~mm, while the main -- linear -- dielectric substrate (namely a Rogers RO3010), of width $h_1-h_2=1.268$~mm, has a relative dielectric constant $\epsilon_{r1}=10.2$. The values for the SRR characteristics are also borrowed from Ref.~\cite{Aznar} and are assumed to take the following values. The internal radius is $r=2.4$~mm, the distance between the rings is $d=0.2$~mm and the rings width is $w=0.6$~mm. As far as the nonlinear dielectric film is concerned, following Ref.~\cite{cai}, we have assumed a strontium barium titanate (SBTO) paraelectric thin film, of width $h_2=2~\mu$m, and  relative dielectric constant $\epsilon_{r2}=300$. Using the above parameter values, we may determine (as per the relevant equations provided in Appendix A) the values of the effective capacitances and inductances involved in the SRR-based CPW structure. This way, we find that $L_{R}=4.11$~nH, $L=0.9$~nH, $L_{s}=1.33$~nH, $C_{s}=4.9$~pF, and $C_{0}=2.44$~pF. Notice that the value of $C_0$ is also obtained consistently from the effective voltage-dependent capacitance of Ref.~\cite{cai}, namely:
\begin{equation}
C(V) = C_{\rm o} \left[1+ \frac{1}{(b_0 + b_1 V_B) +b_1 V} \right],
\label{cainonlin}
\end{equation}
where $C_{\rm o}=1.5$~pF, $b_0 =0.49$, $b_1= 0.25$~V$^{-1}$, while $V_B=4.38$~V is the constant (DC) bias voltage. According to the above expression, the values of the constant coefficients in Eq.~(\ref{eq:Taylor1}) are given by
$C_{0}=C_{\rm o}[1+(b_0+2b_1 V_B)^{-1}] =2.44$~pF [identical to the result of Eq.~(\ref{cr})], and
$C_{0}^{'} = C_{\rm o} b_1 (b_0+2b_1 V_B)^{-2}=0.147$~pF/V.

According to the above, the frequencies in Eq.~(\ref{eq:freq}), which are in the microwave regime, take the values $f_{se}=2.272$~GHz, $f_{sh}=3.395$~GHz, $f_{s}=1.975$~GHz, and $f_{RH}=1.592$~GHz; accordingly, the values of the normalized parameters $\gamma$, $\mu $ and $\beta$ appearing in Eq.~(\ref{eq:model}) take the following values:
\begin{equation}
\gamma=0.66, \qquad \mu=0.58, \qquad \beta=0.47.
\label{param}
\end{equation}
Thus, in our simulations (see Sec.~III.A), we will use the above values to investigate quasi-discrete microwave solitons in this setting, and also discuss possible modifications of the considered setup in order to study how relevant changes in parameter values affect the domains of existence of these nonlinear structures (see Sec.~III.B). Notice that for the above mentioned choice of the physical parameters, the time unit associated to Eq.~(\ref{eq:model}) is $t_0 =(2\pi f_{sh})^{-1} \approx 50$~picoseconds, while the voltage unit is $\upsilon_0 = 2C_{0}/C_{0}^{'} \approx 33$~Volts.

Before proceeding further, we should note the following. As indicated by Eq.~(\ref{eq:Taylor1}), in our considerations we take into regard only the first order approximation in the $C(V_n)$ dependence, while this restriction may not be accurate enough for artificial nonlinearities (induced, e.g., by inserting nonlinear elements, such as diodes, in the SRR slots -- see Ref.~\cite{lapine2} and \cite{lars} for relevant theoretical and experimental studies, respectively). Nevertheless, in our case, where the nonlinearity is induced by the insertion of the nonlinear dielectric thin film (see Fig.~\ref{cpwsrr}), the considered approximation is quite reasonable: indeed, in our numerical simulations (see Sec.~III) we use a value for the initial voltage equal to $V_0 = 0.5$~V (in physical units). For such a value of the voltage (similar, and even smaller, values have also been used in relevant experimental works \cite{koz2,koz3,revtl,lars,ogas,wang}), it can be found that $C(V_0) = C(V_B) + C_{0}^{'} V_0 = (2.44 + 0.0735)~{\rm pF} = 2.51$~pF, i.e., $\approx 3\%$ higher than the value of $C_0 = 2.44$~pF. If we had taken into regard the quadratic term in the Taylor expansion in Eq.~(\ref{eq:Taylor1}), namely $\frac{1}{2}C_{0}^{''}V_n^{2}$, this term would take the value $0.00575$~pF, i.e., only $\approx 0.2\%$ higher than the above mentioned value of $C(V_0) = 2.51$~pF (that corresponds to the lowest order of approximation). Thus, according to the above arguments, and given that the initial voltage value $V_0$ can also be controlled by other additional parameters stemming from our analysis [see parameters $\epsilon$ and $\eta$ in Eq.~(\ref{V0}) below], we will proceed by analyzing Eq.~(\ref{eq:model}) that takes into account only the first order approximation in the $C(V_n)$ dependence
[as per Eq.~(\ref{eq:Taylor1})].

\subsection{The quasi-discrete approximation and the NLS model}

In this Section, we will employ the quasi-discrete approximation (see, e.g., Refs.~\cite{Marquie,Remoissenet}
and Appendix B). Generally, this approach is a variant of the multi-scale perturbation method, which is a well-known powerful tool to derive effective evolution equations (valid under certain conditions and for appropriate spatial and temporal scales) that are much simpler than the original models \cite{Jeffrey}. In our case, since our original model [cf. Eq.~(\ref{eq:model})] is actually a nonlinear dynamical lattice, we adopt the quasi-discrete approximation due to the fact that it takes into regard the discreteness of the system: this approach, allows for the description of quasi-discrete envelope solitons (satisfying an effective NLS model), which are characterized by a discrete carrier and a slowly-varying continuum pulse-like envelope. Notice that, alternatively, one could adopt a continuum approximation, i.e., take the continuum limit of Eq.~(\ref{eq:model}) and analyze the latter in the framework of a multi-scale perturbation scheme (as in Refs.~\cite{nar,ogas,wang}). However, as we will show below (and as was the case in Ref.~\cite{lars}), the quasi-discrete approximation is more accurate than the continuum one on providing estimates for the domains of existence of envelope solitons. Furthermore, the analytically determined soliton profile and characteristics (such as the center of mass and width) will be found to be in good agreement with direct simulations obtained in the framework of the original lattice model of Eq.~(\ref{eq:model}) -- see Sec.~III.A below.

We seek for solutions of Eq.~(\ref{eq:model}) in the form:
\begin{equation}
V_n =\sum_{\ell=1} \epsilon^{\ell} V_{\ell}(X,T)e^{i\ell \theta_n} +{\rm c.c.},
\label{eq:ansatz}
\end{equation}
where $V_{\ell}$ ($\ell=1,2,\cdots$) are unknown envelope functions depending on the slow scales $X=\epsilon (n-v_g t)$ (where $v_g$ is the group velocity, to be determined in a self-consistent manner) and $T = \epsilon^2 t$; here, $0< \epsilon \ll 1$ is a formal small parameter related to the soliton amplitude (see below). Additionally, the function $\exp(i\theta_n)$, with $\theta_n = \omega t - k n$ (with $\omega$ and $k$ denoting frequency and wavenumber, respectively) describes the carrier. In the above ansatz, the envelope (carrier) is obviously continuous (discrete) in space; the results obtained in the framework of the quasi-discrete approximation, may be directly viewed in the continuum limit (and would correspond to the continuum approximation) of $k\rightarrow0$; see also
Appendix C.

Substituting Eq.~(\ref{eq:ansatz}) into Eq.~(\ref{eq:model}) we obtain the following results (see more details in
Appendix B). First, to order $O(\epsilon$) (linear limit), we derive the following dispersion relation:
\begin{equation}
\omega^{4}-(1+\gamma^2+4\beta^2\sin^{2}\frac{k}{2})\omega^{2}+4\beta^2\mu^2\sin^{2}\frac{k}{2}+\gamma^2=0.
\label{eq:dispqc}
\end{equation}
\begin{figure}[tbp]
\centering
\includegraphics[width=7cm]{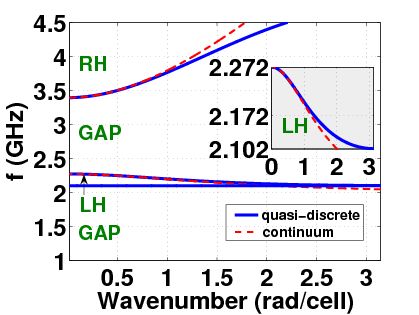}
\caption{(Color online) The dispersion relation, showing the frequency $f$ (in GHz) as a function of the wavenumber $k$ (in rad/cell). Solid (blue) lines and dashed (red) lines correspond to the quasi-discrete and continuum approximations [cf.~Eqs.~(\ref{eq:dispqc}) and (\ref{eq:dispc}), respectively]. There exist a high-frequency RH band, for $3.395$~GHz~$<f<4.734$~GHz, and a low-frequency LH band, for $2.102$~GHz~$<f<2.272$~GHz; the inset shows a magnification of the dispersion relation in the LH frequency band.}
\label{fig:dispersion}
\end{figure}

In Fig.~\ref{fig:dispersion} we plot the frequency $f \equiv \omega/2\pi f_{sh}$ (in GHz) as a function of the  wavenumber $k$ (in rad/cell). It is clear that, apart from the shown gaps (where EM wave propagation is not supported), there exist two different frequency bands (where propagation of EM waves is possible): a high-frequency (HF) and a low-frequency (LF) band, with $3.395$~GHz~$<f<4.734$~GHz and $2.102$~GHz~$<f<2.272$~GHz, respectively. It should be noticed that the lower (upper) cutoff frequency of the HF (LF) frequency band is equal to $f_{sh}$ ($f_{se}$) -- see Eq.~(\ref{eq:freq}). Obviously, in the HF (LF) band the product of the wavenumber $k$ and the group velocity $v_g \equiv \partial \omega/\partial k$ is positive (negative) and, thus, energy and wavefronts travel in the same (opposite) directions in the HF (LF) frequency region. Thus, in the LF band (which is clearly a LH frequency band) the considered SRR-CPW structure apparently behaves as a LH transmission line; the dispersion relation in this LH frequency band, is shown in the inset of Fig.~\ref{fig:dispersion}. Note that in the continuum limit (see pertinent results, for $k \rightarrow 0$, in Appendix C) the LH frequency band under consideration becomes slightly wider, i.e., it extends in the interval $1.976$~GHz~$<f<2.272$~GHz; a similar effect is also observed for the RH band ($f>3.395$~GHz).

Next, proceeding to the next order [$O(\epsilon ^{2})$] in the perturbation scheme, we obtain the group velocity,
given by:
\begin{equation}
v_{g}=\frac{\partial \omega}{\partial k} = \frac{\beta^2\sin k(\omega^{2}-\mu^{2})}{2\omega^{3}-(1+\gamma^{2}+4\beta^2\sin^{2}\frac{k}{2})\omega}.
\label{eq:gvelqc}
\end{equation}
Finally, to order $O(\epsilon ^{3})$, we obtain a nonlinear evolution equation for the unknown voltage $V_1(X,T)$, namely the following NLS equation,
\begin{equation}
i \partial_T V_1+ P \partial_{X}^2 V_1 + Q |V_{1}|^2 V_{1}=0,
\label{eq:NLS}
\end{equation}
with dispersion and nonlinearity coefficients, $P$ and $Q$ respectively, given by the following expressions:
\begin{eqnarray}
P&=&\frac{\partial^{2}\omega}{\partial k^{2}}=
\frac{(1+\gamma^{2}+4\beta^2\sin^{2}\frac{k}{2}-6\omega^{2})v_{g}^{2}}{-4\omega^{3}
+2(1+\gamma^{2}+4\beta^2\sin^{2}\frac{k}{2})\omega} \nonumber\\
&+&
\frac{4\beta^2v_{g}\omega\sin k + \beta^2(\omega^{2}-\mu^{2})\cos k}{-4\omega^{3}+2(1+\gamma^{2}+4\beta^2\sin^{2}\frac{k}{2})\omega},
\label{eq:dispcoef}
\end{eqnarray}
and
\begin{equation}
Q=\frac{4\omega^{4}(\gamma^{2}-\omega^{2})(4\omega^{2}-\gamma^{2})}{[-2\omega^{3}+(1+\gamma^{2}
+4\beta^2\sin^{2}\frac{k}{2})\omega]\mathcal{G}},
\label{eq:nlcoef}
\end{equation}
where the function $\mathcal{G}=\mathcal{G}(\omega,k)$ is given by:
\begin{equation}
\mathcal{G}=16\omega^{4}-4(1+\gamma^{2})\omega^{2}+\gamma^{2}-4\beta^2(4\omega^{2}-\mu^{2})\sin^{2}k.
\label{eq:Gparam1}
\end{equation}

\begin{figure}[tbp]
\begin{tabular}{cc}
\includegraphics[width=7cm]{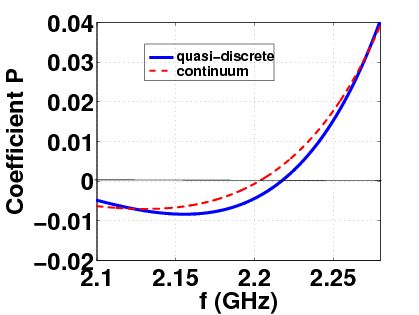} \\
\includegraphics[width=7cm]{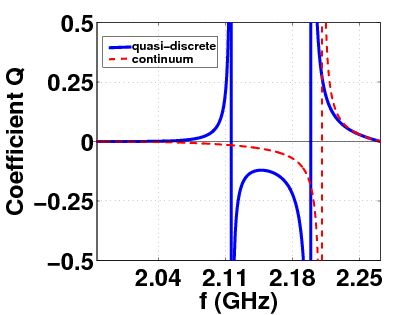}
\end{tabular}
\caption{(Color online) The dispersion coefficient $P$ (top panel) and the nonlinearity coefficient $Q$ (bottom panel) as functions of the frequency $f$ (in GHz) in the LH band. Solid (blue) and dashed (red) lines correspond to the quasi-discrete and continuum approximations, respectively. Note that there exist three characteristic frequencies, namely a frequency where the dispersion coefficient $P$ vanishes, and two frequencies corresponding to poles of the nonlinearity coefficient $Q$; on the contrary, in the continuum approximation, the coefficient $Q$ possesses a sole pole.}
\label{fig:Pcoeff}
\end{figure}

%%%%%%%%%%%%%%%%%%%%%%%%%%%%%%%%%%%%%%%%%%%%%%%%%%%%%%%%%%%%%%%%%%%%%%%%%%%%%%%%%%%%

\subsection{The quasi-discrete soliton solution}

\begin{figure}[tbp]
\centering
\includegraphics[width=7cm]{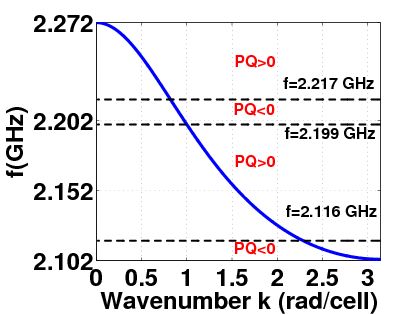}
\caption{(Color online) The signs of the product of the dispersion and nonlinearity coefficients, $P$ and $Q$, for various subintervals of the LH frequency band within the framework of the quasi-discrete approximation. The solid (blue) curve corresponds to the dispersion relation (see inset in Fig.~\ref{fig:dispersion}); the dashed (black) lines at $f=2.116$~GHz and $f=2.199$~GHz indicate the location of the poles of the coefficient $Q$ (see bottom panel of Fig.~\ref{fig:Pcoeff}), while the one at $f=2.217$~GHz shows where the coefficient $P$ vanishes (see top panel of Fig.~\ref{fig:Pcoeff}). Notice that for $PQ>0$ ($PQ<0$) the NLS model supports bright (dark) soliton solutions.}
\label{pr}
\end{figure}

As is well known (see, e.g., Ref.~\cite{yuri}), the NLS equation possesses soliton solutions, the type of which is governed by the signs of the dispersion and nonlinearity coefficients. In particular, if $PQ>0$, the NLS model supports {\it bright} soliton solutions, while for $PQ<0$ it supports {\it dark} soliton solutions. In the case under consideration, the signs of the coefficients $P$ and $Q$ for Eq.~(\ref{eq:NLS}) depend on the frequency. In Fig.~\ref{fig:Pcoeff} we show this frequency dependence of $P$ and $Q$ in the LH frequency band (the figure shows $P$ and $Q$ both in the quasi-discrete and the continuum approximation, corresponding to solid and dashed lines, respectively). First, it can readily be observed that, in the quasi-discrete approximation, the dispersion coefficient (see top panel of Fig.~\ref{fig:Pcoeff}) satisfies $P<0$ ($P>0$) for $f<2.217$~GHz ($f>2.217$~GHz) inside the LH frequency band. On the other hand, the nonlinearity coefficient $Q$ (see bottom panel of Fig.~\ref{fig:Pcoeff}), which has two poles (for $f=2.116$~GHz and $f=2.199$~GHz) inside the LH frequency band is $Q>0$ for $f\in[2.102,2.116)\cup(2.217,2.272]$ and $Q<0$ for $f\in(2.116,2.199)$ (in GHz). Thus, the product $PQ$ takes the following signs: $PQ>0$ for $f\in(2.116,2.199)\cup(2.217,2.272]$, while $PQ<0$ for $f\in[2.102,2.116)\cup(2.199,2.217)$ (in GHz); the above results are summarized and demonstrated in Fig.~\ref{pr}.

Here, it should also be noticed that in the continuum approximation the nonlinearity coefficient $Q$ has only one pole (at $f=2.21$~GHz); as a result, this approximation estimates a broader interval wherein bright solitons can be formed: as may be seen in Fig.~\ref{fig:Pcoeff} (see also Appendix C), bright solitons can be formed in the intervals $f\in[1.976,2.203)\cup(2.21,2.272]$ (in GHz). While this pole arises close to the point where a pole emerges in the quasi-discrete approximation, the other pole of the latter seems to be missed by the presumably less accurate  genuinely continuum approximation. In any case, the above investigation shows that bright solitons (corresponding to $PQ>0$) are easier to be observed in the SRR-CPW structure: first, unlike dark solitons (corresponding to $PQ<0$), bright solitons are supported in a wide subinterval of frequencies inside the LH frequency band; on the other hand, they can be formed for carrier frequencies sufficiently far away from frequencies where the dispersion coefficient $P$ vanishes or the nonlinearity coefficient $Q$ has resonances (see Fig.~\ref{fig:Pcoeff}). This is where we expect this type of approximations to be most relevant in describing the fully discrete dynamical system. Thus, below, we will confine our considerations to the case of bright soliton solutions of Eq.~(\ref{eq:NLS}); a single soliton solution can be expressed in the following form (see, e.g., Ref.~\cite{yuri}):
\begin{equation}
V_1(X,T) = \eta {\rm sech}[\eta(X-X_0(t))]\exp[i(KX-\Omega T)],
\label{sol1}
\end{equation}
where $\eta$ is the soliton's amplitude (and inverse width), $X_0(t)\equiv X_0(0)+VT$ is the soliton center, $X_0(0)$ is the initial soliton location, $V$ is the soliton velocity, $K=V$ denotes the soliton's wavenumber, while $\Omega=(1/2)(K^2-\eta^2)$ is the soliton's frequency. The above expression can be used to approximate the unknown voltage $V_n(t)$ in Eq.~(\ref{eq:model}), in terms of the original coordinates, as follows:
\begin{eqnarray}
V_n(t) \approx V_{0}\sech[\epsilon\eta(n-c_0 t-n_0)] \cos(K_0 n-\Omega_0t),
\label{eq:brightsolution}
\end{eqnarray}
where $n_0$ is the initial soliton position, while the soliton's amplitude $V_0$, velocity $c_0$, wavenumber $K_0$, and frequency $\Omega_0$, are respectively given by:
\begin{eqnarray}
V_{0}&=&2\epsilon\eta \sqrt{2|PQ^{-1}|},
\label{V0} \\
c_0 &=& v_g+2\epsilon K P,
\label{C0} \\
K_0 &=& k+\epsilon K,
\label{K0} \\
\Omega_0 &=& \omega+\epsilon (Kv_{g}+2\epsilon\Omega P).
\label{Om}
\end{eqnarray}

We conclude this Section by mentioning the following. The presented approximate soliton solutions in the LH frequency band have a unique feature as compared to ones that can be formed in the RH frequency band (i.e., for $3.395$~GHz~$<f<4.734$~GHz -- see Fig.~\ref{fig:dispersion}). This can be understood by the fact that while in the RH regime the frequency increases with the wavenumber, in the LH regime the transmission line exhibits ``anomalous'' dispersion, and the frequency decreases with the wave number, as seen in Fig.~\ref{fig:dispersion}. Thus, the solitons in the LH frequency band are actually {\it backward waves}, with group and phase velocities that are antiparallel to each other (see also discussion and relevant experimental observations in Refs.~\cite{revtl,lars}).

\section{Numerical results}

\subsection{Analysis of the ``regular'' SRR-CPW structure}

Let us now proceed to study numerically the evolution of the quasi-discrete solitons presented in the previous section in the framework of the fully discrete model of Eq.~(\ref{eq:model}). First, we will study the latter, taking parameter values as per Eq.~(\ref{param}) (recall that these were borrowed from Ref.~\cite{Aznar}, which reported realization of this -- as characterized here -- ``regular'' SRR-CPW structure) and in the next Section, we will study experimentally relevant modifications in the SRR-CPW structure, so as to investigate how pertinent parameter changes affect the domains of existence of quasi-discrete solitons.

In the results below, we have fixed the parameters related to the soliton's amplitude as $\eta=1$ and $\varepsilon=0.02$, which correspond to an initial value of the voltage equal to $V_0=0.5$~V (similar, and even smaller, values have also been used in experiments \cite{koz2,koz3,revtl,lars,ogas,wang}). Furthermore, we have fixed the initial soliton position to $X_0(0)=1/2$, and we have varied the frequency $f$ and the soliton wavenumber $K$ (recall that the latter sets the initial soliton momentum). The chosen intervals of variation have been $f\in[2.11,2.18]\cup[2.21,2.25]$ (in GHz) and $K \in[0,\pi]$. Those values of $f$ lie in the LH regime with both $P<0$ and $Q<0$. A ring of $N=1001$ cells (and periodic boundary conditions) has been chosen for the simulations.

In order to characterize the outcome of the simulations, and compare analytical and numerical results, we have defined two diagnostics. The first one is the evolution of the center (alias pseudo-center-of-mass),
\begin{equation}
X(t)=\frac{\sum_n n V_n^2}{\sum_n V_n^2},
\label{X}
\end{equation}
and the second one is a measure of the width (alias pseudo-width), defined as
\begin{equation}
W(t)=\sqrt{\frac{\sum_n n^2 V_n^2}{\sum_n V_n^2}-X^2(t)}.
\label{W}
\end{equation}
According to the results of the previous Section, analytical forms for these quantities can readily be found in the continuous setting [i.e., approximating the soliton as per Eq.~(\ref{sol1})]:
\begin{eqnarray}
X(t)&=&X_0(0)+c_0 t = \frac{1}{2}+c_0 t, \\
\label{Xc}
W(t)&=&
%\frac{1}{4}+\frac{\pi^2}{12\epsilon^2\eta^2}+C_0 t+C_0^2 t^2=
\frac{\pi}{2\sqrt{3}\epsilon\eta}.
\label{Wc}
\end{eqnarray}

\begin{figure}
    \begin{tabular}{cc}
    \includegraphics[width=3.85cm,height=3.14cm]{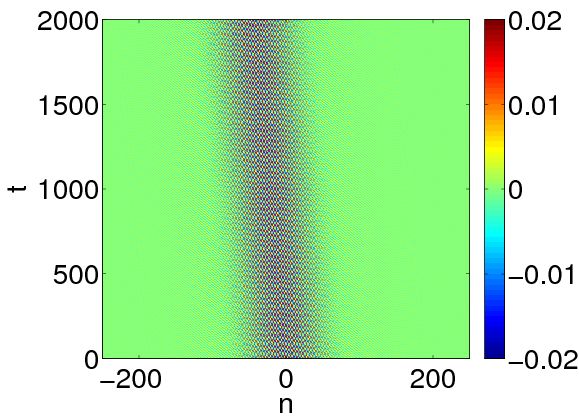} &
    \includegraphics[width=3.85cm]{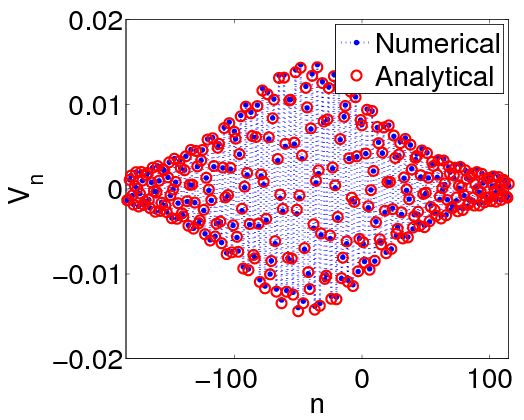} \\
    \includegraphics[width=3.85cm]{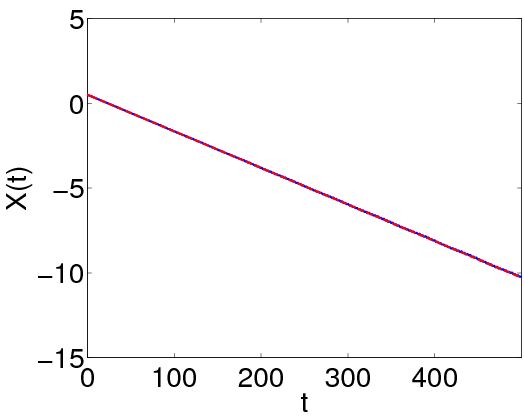} &
    \includegraphics[width=3.85cm]{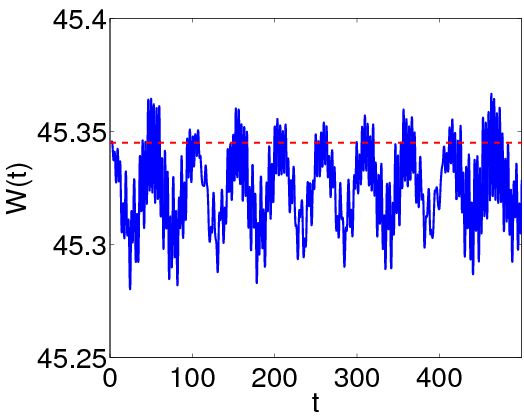} \\
    \end{tabular}
\caption{(Color online) Top left: density plot of the time evolution of $V_n$ obtained numerically. The top right panel compares the analytical and numerical profiles of $V_n$ at $t=2000$. The bottom panels show the time evolution of the center of mass (left) and the width diagnostic (right); in both cases, the solid line corresponds to the numerics and the dashed line to the analytical prediction. The parameters used are $f=2.15$~GHz and $K=0$.}
\label{fig:simul1}
\end{figure}

\begin{figure}
    \begin{tabular}{cc}
    \includegraphics[width=3.85cm,height=3.14cm]{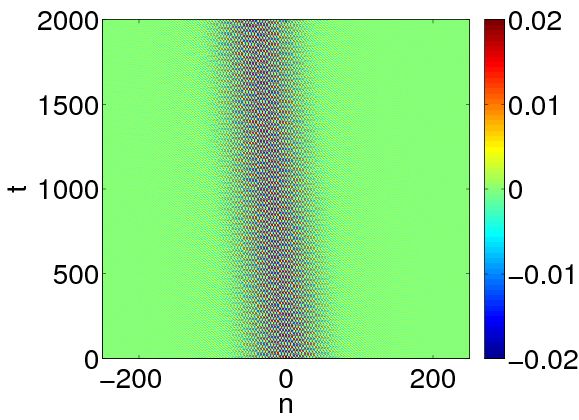} &
    \includegraphics[width=3.85cm]{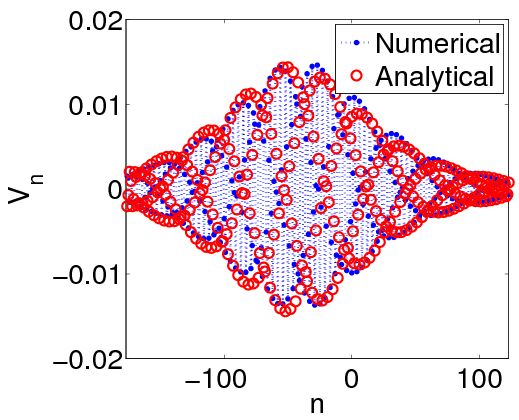} \\
    \includegraphics[width=3.85cm]{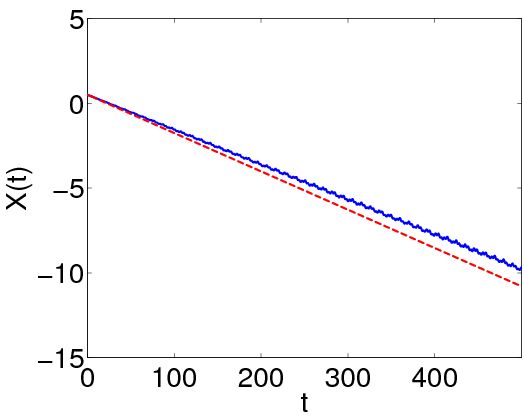} &
    \includegraphics[width=3.85cm]{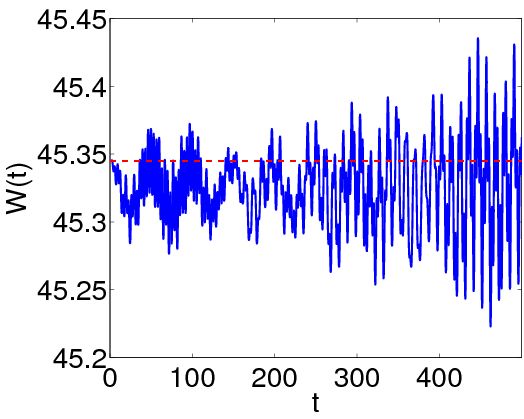} \\
    \end{tabular}
\caption{(Color online) Same as Fig.~\ref{fig:simul1} but for $f=2.15$ GHz and $K=\pi$.}\label{fig:simul2}
\end{figure}

\begin{figure}
    \begin{tabular}{cc}
    \includegraphics[width=3.85cm,height=3.14cm]{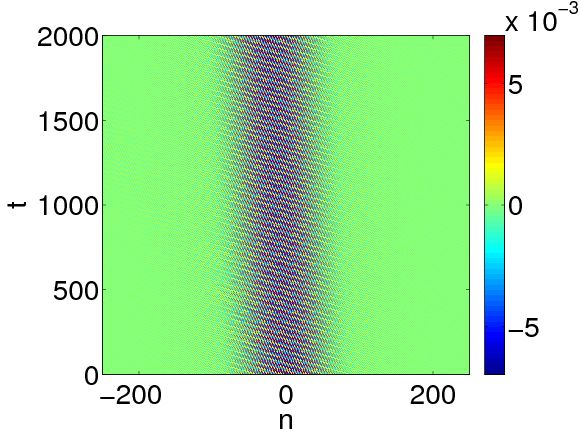} &
    \includegraphics[width=3.85cm]{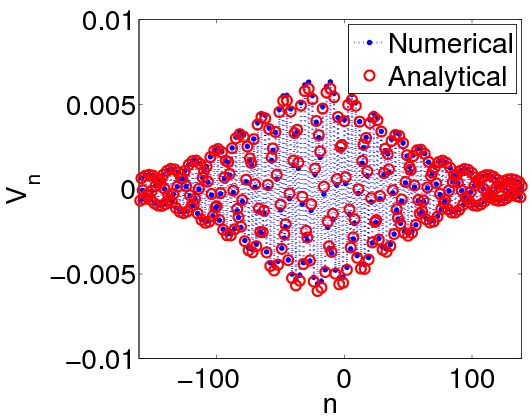} \\
    \includegraphics[width=3.85cm]{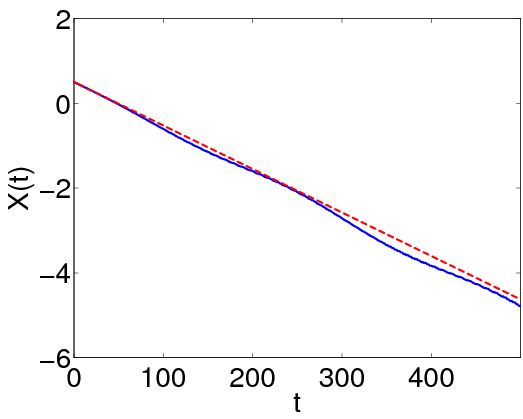} &
    \includegraphics[width=3.85cm]{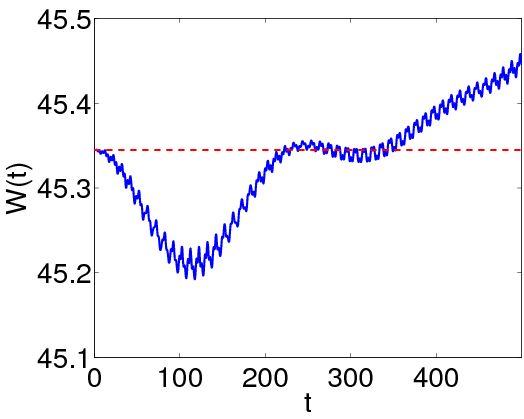} \\
    \end{tabular}
\caption{(Color online) Same as Fig.~\ref{fig:simul1} but for $f=2.11$ GHz and $K=0$.}\label{fig:simul3}
\end{figure}

\begin{figure}
    \begin{tabular}{cc}
    \includegraphics[width=3.85cm,height=3.14cm]{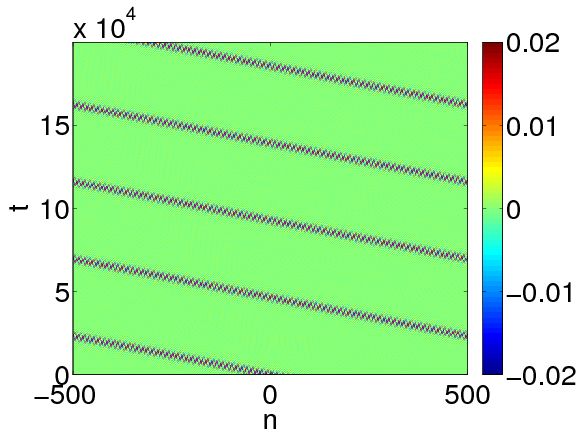} &
    \includegraphics[width=3.85cm]{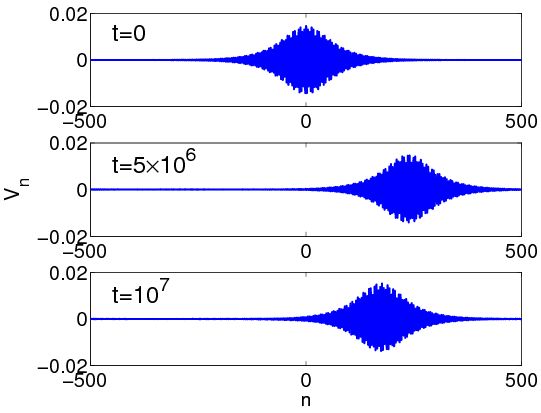} \\
    \includegraphics[width=3.85cm]{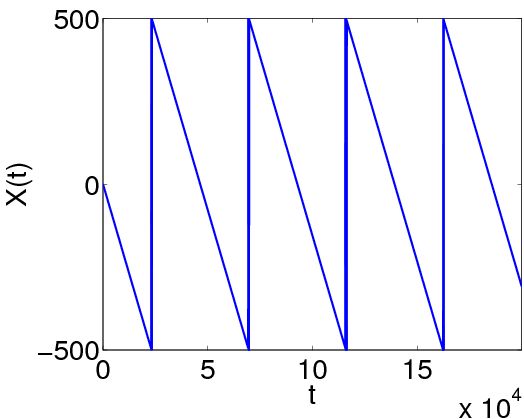} &
    \includegraphics[width=3.85cm]{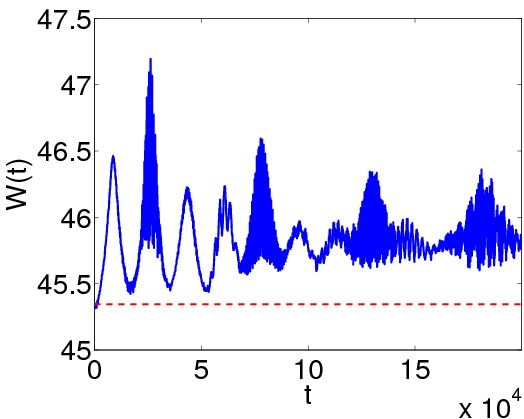} \\
    \end{tabular}
\caption{(Color online) The soliton of Fig. \ref{fig:simul1} is evolved until $t=2\times10^5$. All the panels are similar to that of Fig. \ref{fig:simul1} except for the top right. In this panel, snapshots of the soliton at $t=2\times10^6$ and $t=10^7$ are compared to the initial condition of the simulation in order
to examine its robustness under a very lengthy time evolution.}
\label{fig:simullong}
\end{figure}

Below we provide the outcome of some typical simulations (see Figs.~\ref{fig:simul1}-\ref{fig:simullong}) through density plots of $V_n$, the spatial profile of $V_n$ at $t=2000$, as well as the time evolution of the center of mass $X(t)$ and the width parameter $W(t)$. Generally, as we show below in more detail, the direct numerical integration of Eq.~(\ref{eq:model}), with initial conditions borrowed from the analytical expression of Eq.~(\ref{eq:brightsolution}) (for $t=0$), have revealed the following: the quasi-discrete bright solitons exist, indeed, in the predicted frequency regions inside the LH frequency band; furthermore, their form, as well as the evolution of their center and width, can be well approximated by pertinent analytical expressions provided above, especially in cases where the carrier frequency is chosen sufficiently different from certain characteristic frequencies (i.e., where the dispersion coefficient $P$ vanishes or the nonlinearity coefficient $Q$ has resonances).
%, and (b) when the initial soliton momentum is $K=0$.
%The latter condition (namely the zero initial momentum) can be physically understood by the following fact: when discrete solitons are initially forced to move, they have to jump from site to site, passing -- typically -- from stable to unstable configurations (see, e.g., Ref.~\cite{yuri}) and, thus, the soliton mobility is generally restricted on the lattice. Accordingly, due to their restricted mobility, our quasi-discrete solitons can not move similarly to their continuum counterparts obeying the effective NLS Eq.~(\ref{eq:NLS}).
%

Before proceeding with the description of our results, we should also note the following. Although most of our simulations were performed for relatively large normalized times -- typically of order of $t\sim 10^3$ -- given our time normalization, the physical unit time (set by the frequency $f_{sh}=3.39$~GHz) is very small, namely $t_0 =(2\pi f_{sh})^{-1} \approx 50$~picoseconds (see Sec.~II.A). In fact, since all characteristic frequencies of the system (see Eq.~(\ref{eq:freq})) are in the microwave regime, all characteristic times are less than a nanosecond, rendering long simulations extremely time-consuming. Nevertheless, in a particular case where the condition above is fulfilled, we have performed a few extremely long simulations (with normalized time horizons of $t=10^7$, corresponding to a physical time of the order of a millisecond), finding that the agreement reported below is still upheld in these runs. This indicates that our predictions concerning soliton formation and robustness may be valid for experimentally relevant times.

Let us expose our results starting with Fig.~\ref{fig:simul1}, which shows the case of a quasi-discrete soliton with carrier frequency $f=2.18$~GHz and zero initial momentum, $K=0$, which evolves as a stable object over long times. In this case, the agreement between analytical and numerical results pertaining to the soliton profile, but also to the evolution of the center of mass and width diagnostics, is very good. On the other hand, Fig.~\ref{fig:simul2} shows the evolution of a soliton with also $f=2.18$ GHz but with nonzero soliton momentum, $K=\pi$ (with similar conclusions), while the soliton of Fig.~\ref{fig:simul3} corresponds to $f=2.11$ GHz (for $K=0$), which is close to the resonance of the nonlinearity coefficient $Q$. In this last case, it is clear that although quasi-discrete bright solitons exist, the agreement between analytics and numerics becomes worse (especially as concerns the estimation of the soliton width parameter shown in the bottom right panel of Fig.~\ref{fig:simul3}). This can be attributed to the proximity to the resonance where we expect the conditions for the quasi-discrete approximation to be violated.
%Note that a similar situation is also observed in the case of Fig.~\ref{fig:simul4}, which shows a quasi-discrete soliton with carrier frequency lying in the upper (second) allowable regime inside the LH band: in this case, $f=2.23$~GHz and $K=0$.
Note that similar results have also been obtained for the upper (second) allowable regime inside the LH band [$f\in(2.217,2.272]$ (in GHZ)].

Finally, using parameter values corresponding to the case shown in Fig.~\ref{fig:simul1} (i.e., for $f=2.15$~GHz and $K=0$), we have performed a very long simulation, up to normalized times $t=10^7$ (physical time $\sim 1$~ms) in order to check numerically the soliton robustness; details are shown in Fig.~\ref{fig:simullong}.
%It is obvious that the soliton persists as a stable object up to this long simulation time (where the numerical integration was stopped).
As far as the width parameter $W$ is concerned (bottom right panel of the figure), it is clear that -- despite its fluctuations -- it possesses an almost constant mean value which (although not captured precisely by the analytical approximation) indicates that the initial pulse does not spread out. Furthermore, the top panels of the figure -- and particularly the snapshots of the pulse profile at $t = 10^7$ -- clearly indicate that the soliton persists as a stable object up to this long simulation time (where the numerical integration was stopped).
%This particular figure, as well other results presented above,
These results, as well as the ones presented above, indicate that quasi-discrete solitons can be formed in SRR-based CPW nonlinear structures, and may propagate over experimentally relevant times.

\subsection{Modifications of the ``regular'' SRR-CPW structure}

In the previous section, we have studied in detail quasi-discrete solitons of Eq.~(\ref{eq:model}) that can be formed in the SRR-CPW structure for parameter values given in Eq.~(\ref{param}). Here, we will study different scenarios arising from modifications of either the CPW structure or the SRR geometry. As it is clear from Fig.~\ref{cpwsrr} (and also the discussion in Sec.~II.A and Appendix A), there is a considerable degree of flexibility as concerns the choice of the parameter values that are involved in the determination of the parameters $\gamma$, $\mu$ and $\beta$: indeed, one may consider different characteristic widths $h_1$, $h_2$ of the dielectrics (and values of the relative dielectric constants $\epsilon_{r1}$, $\epsilon_{r2}$ thereof), or different geometrical characteristics of the CPW structure or of the SRRs.

Nevertheless, since the considered SRR-CPW structure has already been optimally realized in practice \cite{Martin,Aznar}, one should consider changes that keep the basic characteristics of the configuration as close to its experimental
realization as possible. We thus choose to keep the characteristics of
the ``main'' dielectric substrate (characterized by the parameters $h_1$ and $\epsilon_{r1}$), as well as the transverse width $a$ of the CPW structure fixed.
Furthermore, we also keep fixed the characteristics of the nonlinear dielectric (with parameters $h_2$ and $\epsilon_{r2}$), since -- according to our considerations -- they do not significantly affect the linear response of the system. On the other hand, we consider certain changes that would arise from a slightly different realization of the considered setup: in particular, we will study the changes in the the width $G$ of the slots for the CPW structure, and changes of the geometric characteristics of the SRRs, namely their radius $r$ (for fixed width $w$ and spacing $d$ between the SRRs). As we will show below, these changes may improve the nonlinear SRR-CPW configuration, in the sense that they lead to an increase of the width of the LH frequency band, and also increase the ``central'' frequency band (i.e., in between the resonances of coefficient $Q$) where bright solitons can be formed ($PQ>0$). Other changes in the parameter values have also been studied, but the results will not be exposed here, as they lead to results qualitatively similar to the ones presented above, or even worse (i.e., they lead to decrease of the widths of the LH frequency band and/or the domains of existence of bright solitons).

\begin{figure}[tbp]
\centering
\includegraphics[width=6.7cm]{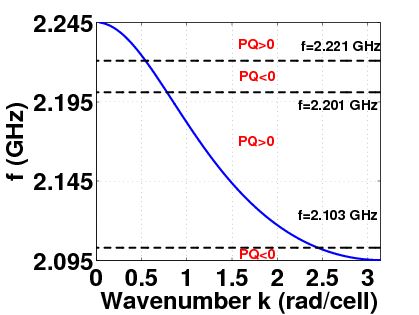}
\includegraphics[width=6.05cm]{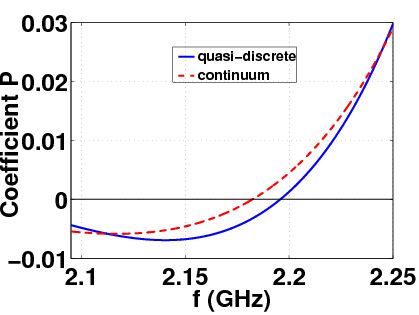}
\includegraphics[width=6.7cm]{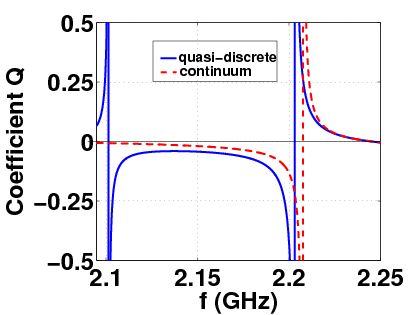}
\caption{(Color online) Top panel: Same as Fig.~\ref{pr}, but for a slot width $G=1$~mm. Middle and bottom panels: Same as Fig.~\ref{fig:Pcoeff} (for $G=1$~mm). In this case, the values of parameters $\gamma$, $\mu$ and $\beta$ are given in Eq.~(\ref{g1}).
}
\label{pr2}
\end{figure}

A relevant study along the above lines has led to the following results. First, for a fixed value of the distance between the slots, $D=7$~mm, we have changed the value of the slot width $G$ in the interval $0.7$~mm -- $1.5$~mm (recall that originally $G=1.35$~mm). Such changes result in different values of $\kappa_j$ ($j=0,1,2$) in Eq.~(\ref{kj}) and, accordingly, to different values of the effective circuit model parameters $C_R$, $C_L$, $L_L$, $C_s$ and $L_s$, which, in turn, provide different values for the parameters $\gamma$, $\mu$ and $\beta$ [cf. Eq.~(\ref{eq:freq})]. As $G$ is decreased (increased) in the aforementioned interval, the parameters $\gamma$ and $\mu$ are also decreased (increased), while the parameter $\beta$ is increased (decreased). These changes result in a decrease (increase) of the LH frequency, but also lead to an increase (decrease) of the ``central'' frequency band where bright solitons can be formed. As an example, in the top panel of Fig.~\ref{pr2}, we show the LH frequency band, as well as the regimes for soliton formation, in a case corresponding to $G=1$~mm, for which the characteristic parameters of the configuration take the values:
\begin{equation}
\gamma=0.61, \qquad \mu=0.54, \qquad \beta=0.48.
\label{g1}
\end{equation}
As seen in this figure, the LH band extends from $2.095$~GHz to $2.245$~GHz (i.e., it is decreased by $\approx 11\%$), while the central frequency region, where bright solitons exist, extends from $2.103$~GHz to $2.201$~GHz (i.e., it is increased by $\approx 15\%$). Notice that the functional form of the parameters $P$ and $Q$ shown in the middle and bottom panels of Fig.~\ref{pr2}, respectively, are similar to the ones in Fig.~\ref{fig:Pcoeff}. One should notice, however, that the parameter $Q$ exhibits an almost flat profile in the central frequency band with $PQ>0$.

\begin{figure}
    \begin{tabular}{cc}
    \includegraphics[width=3.85cm,height=3.14cm]{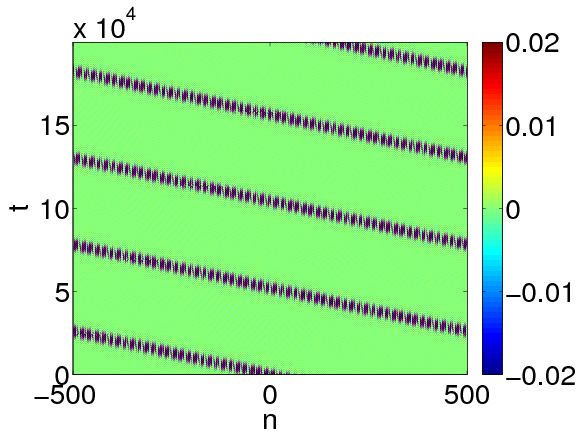} &
    \includegraphics[width=3.85cm]{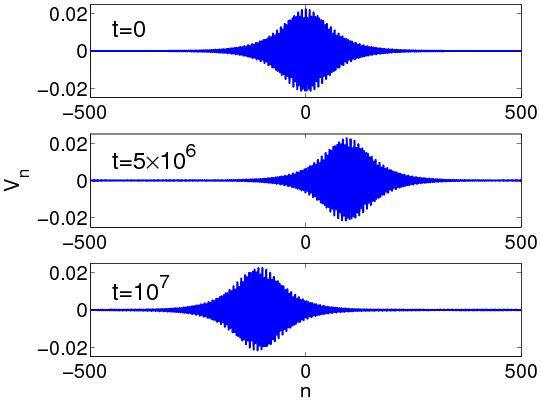} \\
    \includegraphics[width=3.85cm]{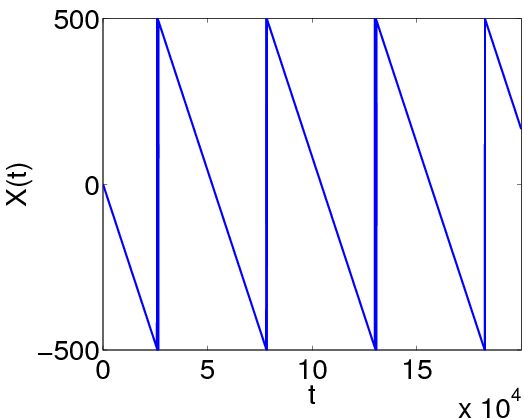} &
    \includegraphics[width=3.85cm]{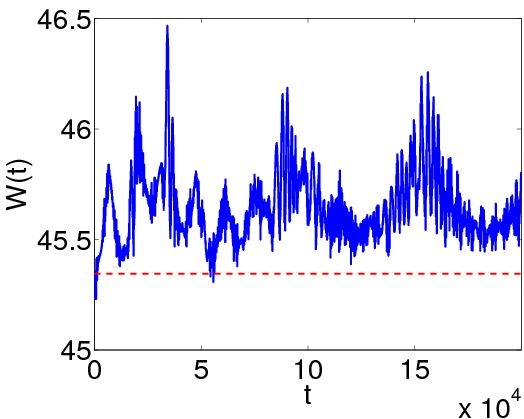} \\
    \end{tabular}
\caption{(Color online) Same as Fig.~\ref{fig:simullong}, but for parameter values given by Eq.~(\ref{g1}). Notice that
the soliton is evolved until $t=2\times10^4$ (top left panel), but it persists for much longer times, namely up to $t=10^7$ (see snapshots in the top right panels). Once again, the width parameter $W$ (bottom right panel) possesses an almost constant value, which indicates the soliton-like nature of the initial pulse.}
\label{slong2}
\end{figure}

Numerical simulations shown in Fig.~\ref{slong2} for bright soliton propagation at the frequency $f=2.15$~GHz lead to results qualitatively similar to the ones presented above (cf. Figs.~\ref{fig:simul1} and \ref{fig:simullong}). In particular, once again, the soliton is quite robust up to long times (see the snapshots corresponding to $t=5\times10^6$ and $t=10^7$ in the top right panel of Fig.~\ref{slong2}), while the width parameter $W$ (bottom right panel of Fig.~\ref{slong2}) has an almost constant mean value (close to the value corresponding to the analytical estimate) indicates that the initial pulse does not spread out, thus featuring genuine soliton characteristics.

%\begin{figure}[tbp]
%\centering
%\includegraphics[width=6.7cm]{LHr26.jpg}
%\includegraphics[width=6.22cm]{PCoeffr26.jpg}
%\includegraphics[width=6.22cm]{QCoeffr26.jpg}
%\caption{(Color online) Same as Fig.~\ref{pr2}, but for a ring radius $r=2.6$~mm. In this case, the values of parameters $\gamma$, $\mu$ and $\beta$ are given in Eq.~(\ref{r1}). }
%\label{pr3}
%\end{figure}

Next, keeping fixed the parameter values of the CPW structure as in Sec.~II.A (in this case $G=1.35$~mm), we have studied changes in the SRR geometry, namely the radius $r$ and width $w$ of the SRRs (for fixed spacing between the SRRs, $d=0.2$~mm). First, in the case with fixed $w=0.6$~mm, the increase (decrease) of the SRR radius $r$ -- in the interval $1.8$~mm to $1.6$~mm -- results in a decrease (increase) of the values of parameters $\gamma$ and $\mu$, while the parameter $\beta$ takes the approximately constant value of $\beta \approx 0.48$. Accordingly, it is found that the width of the LH frequency band is increased (decreased), while the central frequency regime where bright solitons can be formed is increased (decreased) as well. As an example, for $r=2.6$~mm (recall that $r=2.4$~mm for the ``regular'' SRR-CPW structure), the LH frequency band extends from $2.106$~GHz up to $2.343$~GHz, i.e., it has increased $\approx 39\%$.
%-- see top panel of Fig.~\ref{pr3}.
Furthermore, in the same case, the central frequency region where bright
solitons can be formed is also increased by $\approx 29\%$,
%(see top panel of Fig.~\ref{pr3}),
with the frequency dependence of coefficient $Q$
%(see bottom panel of Fig.~\ref{pr3})
being similar to the one shown in the bottom panel of Fig.~\ref{pr2}.
In this case, the parameters $\gamma$, $\mu$ and $\beta$ take the
following values:
\begin{equation}
\gamma=0.69, \qquad \mu=0.57, \qquad \beta=0.48.
\label{r1}
\end{equation}
Numerical simulations
%, shown in Fig.~\ref{slong3},
for bright soliton propagation in this regime
(e.g., we have performed long evolution runs for the
frequency $f=2.181$~GHz) lead to results qualitatively similar to the ones
presented above (cf. Figs.~\ref{fig:simul1}, \ref{fig:simullong} and \ref{slong2}).

%\begin{figure}
%    \begin{tabular}{cc}
%    \includegraphics[width=3.85cm,height=3.14cm]{figadd2a.jpg} &
%    \includegraphics[width=3.85cm]{figadd2b.jpg} \\
%    \includegraphics[width=3.85cm]{figadd2c.jpg} &
%    \includegraphics[width=3.85cm]{figadd2d.jpg} \\
%    \end{tabular}
%\caption{(Color online) Same as Fig.~\ref{slong2}, but for parameter values given by Eq.~(\ref{r1}).}
%\label{slong3}
%\end{figure}

%%%%%%%%%%%%%%%%%%%%%%%%%%%%%%%%%%%

\section{Discussion and Conclusions}

We have studied analytically and numerically quasi-discrete microwave bright solitons that can be formed in the left-handed frequency band of a split ring resonator (SRR) based coplanar waveguide (CPW) structure. We have used the nonlinear transmission line analogue of this structure to derive a nonlinear lattice equation governing the voltage across the fundamental (unit cell) element of the system. This lattice equation was then treated analytically, by means of the quasi-discrete approximation. The latter is a variant of the multi-scale perturbation method, which takes into regard the discreteness of the system by considering the carrier (envelope) of the wave as a discrete (continuum) object. This approach allowed us to derive, in the small-amplitude approximation and for certain space- and time-scales, a nonlinear Schr\"{o}dinger (NLS) model for the unknown voltage envelope function. The NLS model was then used to predict formation of bright solitons in certain sub-regions of the left-handed frequency band. Importantly, the conditions for soliton formation were found to depend on the {\it discreteness} of the system: in fact, if the continuum -- instead of the quasi-discrete -- approximation was used, then the allowable (for soliton formation) frequency bands would be significantly modified. Furthermore, the quasi-discrete approximation predicts effects, such as the appearance of resonance frequencies in the coefficients of the effective NLS model, which suggest optimum operating frequency bands for the observation of quasi-discrete solitons. Generalizing these observations, one should expect that the adopted analytical approach, based on the quasi-discrete approximation, should provide more reliable results concerning conditions for soliton formation in left-handed structures.

Numerical simulations performed in the framework of the nonlinear lattice equation, with initial conditions borrowed from the effective NLS equation, revealed that the (bright) solitons may indeed be formed in the frequency sub-intervals predicted in the analytics. Furthermore, it was shown that if the carrier frequency is chosen to be sufficiently far from characteristic frequencies (where the dispersion and nonlinearity coefficients of the effective NLS model vanish or have resonance poles), the numerically found soliton profile and characteristics (center of mass and width) are in very good agreement with the analytical predictions. Although the numerical simulations were performed for relatively large computation times, the corresponding physical times were small due to the fact that all the characteristic frequencies of the system were in the microwave regime: under any choice, the physical time unit is less than a nanosecond and, thus, simulations corresponding to a few seconds would correspond to computation times of order of $10^{10}$. This difficulty, particular to the microwave structure under consideration, did not allow us to be definitively conclusive as concerns the
robustness of the solitons for realistic experimental times. Nevertheless, very long simulations (corresponding to computing times $\sim 10^7$ or $\sim 1$~millisecond in physical units) have shown that the predicted quasi-discrete microwave solitons are quite robust and do not appear to be modified during their propagation. This, in turn, allows us to conjecture that these nonlinear wave structures have a good chance to be observed experimentally in the near future.

We have also presented a study showing how certain physical parameters of the configuration may affect the results. In particular, we have focused on two cases, namely the effects of decrease of the width of the slots in the CPW structure, and of the increase of the SRR radius. These choices led to modified values of the parameters of the model ($\beta$, $\gamma$, and $\mu$), such that the width of the left-handed frequency band or/and the domain of existence of quasi-discrete bright solitons are increased. This way, we have also proposed certain experimentally relevant changes in the considered configuration, to facilitate observation of solitons in future experiments.

There are many interesting directions for future studies. In that regard, first, we should mention that in the analysis of the considered SRR-CPW structure we have actually excluded the SRR from nonlinearity. It would be interesting to see what happens if SRR is also subject to nonlinear modulation: this is certainly a very challenging direction, in terms of the development of both analytical and numerical techniques, due to the form of the pertinent models, which are coupled nonlinear lattice equations [see, e.g., Eqs.~(\ref{eq:mod01})-(\ref{eq:mod02})]. Such studies would lead to potentially interesting and relevant results concerning nonlinear wave propagation in such settings, as well as the design of nonlinear left-handed transmission lines (and related structures).

On the other hand, we note that following the analytical procedure adopted in this work, it would be possible to analyze soliton formation in relevant left-handed structures. Furthermore, employing this analytical approach, one could -- in principle -- derive self-consistently [at $O(\epsilon ^{4})$ in the perturbation scheme] a generalized NLS model, incorporating higher-order dispersive and nonlinear effects (see, e.g., Ref.~\cite{tsitsas1}). Such a higher-order NLS model could also predict soliton propagation but, in this case, the soliton characteristics would be modified by the presence of the higher-order effects. Thus, an interesting challenge would be the derivation of such a higher-order NLS model, the study of its soliton solutions and a comparison of such findings to direct simulations.

Finally, while the theory and simulations presented in this work assume ideal components, for which excellent performance can be observed, the reality to be met in possible experiments might deteriorate the performance by unavoidable dispersion in component parameters. The effect of disorder is expected to be quite significant in nonlinear settings: in fact, its paramount importance has been demonstrated even in the simpler case of linear resonant systems (see, e.g., Ref.~\cite{disorder}) and, thus, the presence of disorder may drastically affect the results in nonlinear structures. In that regard, a study of how much the considered system (or other relevant ones) is sensitive to disorder is certainly a relevant and important direction for future studies.

\section*{Acknowledgments}
J.C. acknowledges financial support from the MICINN project FIS2008-04848. P.G.K. gratefully acknowledges support from the
NSF (DMS-0806762 and CMMI-1000337), the Alexander von Humboldt Foundation and the Alexander S. Onassis Public Benefit Foundation. The work of D.J.F. was partially supported by the Special Account for Research Grants of the University of Athens.

\appendix
\section{The physical parameters of the SRR-CPW structure}

The elements $C_R$, $L_R$ and $L_L$ associated with the CPW structure are connected with the physical parameters of the system by means of the following equations (see, e.g., Ref.~\cite{chenchu}):
\begin{eqnarray}
C_R&=&2\epsilon_0 \frac{K(\kappa_0)}{K'(\kappa_0)}+\epsilon_0 (\epsilon_{r1}-1)\frac{K(\kappa_1)}{K'(\kappa_1)} \nonumber \\
&+& \epsilon_0 (\epsilon_{r2}-\epsilon_{r1})\frac{K(\kappa_2)}{K'(\kappa_2)},
\label{cr}\\
L_R&=&\left(1+\frac{L_{\rm CPW}}{4L_p}\right)L_{\rm CPW}-L_s,
\label{lr}\\
L_L&=&\frac{1}{2} L_{\rm CPW} +2L_p.
\label{ll}
\end{eqnarray}
In the above expressions, $\epsilon_0$ is the dielectric constant of vacuum, $K$ and $K'$ are the complete elliptic integral of the first kind and its complementary function, respectively \cite{abram}, the arguments of these functions are given by:
\begin{eqnarray}
\kappa_0=\frac{D}{D+2G}, \,\,\,\
%\label{k0}\\
\kappa_j=\frac{\sinh\left(\frac{\pi D}{4h_j}\right)}{\sinh\left[\frac{\pi (D+2G)}{4h_j}\right]} \,\,\,\, (j=1,2),
\label{kj}
%\\k_2&=&\frac{\sinh(\frac{\pi W}{4h_2})}{\sinh(\frac{\pi (W+2G)}{4h_2})}\label{k2}
\end{eqnarray}
and, finally, $L_p$ and $L_{\rm CPW} = 4\epsilon_0 (30\pi)^2 K'(\kappa_0)/K(\kappa_0)$ denote, respectively, the inductance of the shunt strip and the effective inductance of the CPW structure (note that the former takes the approximate value $L_p \approx 0.36$~nH \cite{Aznar}). As far as the values of the SRR parameters, $L_{s}$ and $C_{s}$, are concerned, they are given by the following expressions \cite{Aznar}:
\begin{eqnarray}
\!\!\!\!\!\!\!\!\!\!\!
L_s &=& \frac{2F^2 L_{\rm CPW}^2}{L_{\rm SRR}} \frac{(1+L_{\rm CPW}/4L_p)^2}{1+F^2 L_{\rm CPW}^2/2L_p L_{\rm SRR}},
\label{ls} \\
\!\!\!\!\!\!\!\!\!\!\!
C_s &=& \frac{L_{\rm SRR}^2 C_{\rm SRR}}{2F^2 L_{\rm CPW}^2}
\left( \frac{1+F^2 L_{\rm CPW}^2/2L_p L_{\rm SRR}}{1+L_{\rm CPW}/4L_p} \right)^2.
\label{cs}
\end{eqnarray}
In the above expressions, $F \approx 0.54$ is the fractional area of the slots occupied by the rings \cite{Aznar}, while
$L_{\rm SRR}$ and $C_{\rm SRR}$ denote, respectively, the inductance and capacitance of the SRRs, and are given by:
\begin{eqnarray}
&&\frac{4}{\mu_0}L_{\rm SRR} = (r+\frac{w}{2})\ln\left[ \frac{8(2r+w)}{w}-2\right] \nonumber \\
&&+ \left(r+d+\frac{3w}{2}\right)\ln\left[ \frac{8(2r+3w+2d)}{w}-2\right],
\label{lsrr} \\
&&C_{\rm SRR} = 4\epsilon_0 \frac{K(\kappa_3)}{K'(\kappa_3)}+2\epsilon_0 (\epsilon_{r1}-1)\frac{K(\kappa_4)}{K'(\kappa_4)},
\label{csrr}
\end{eqnarray}
where $\mu_0$ is the permeability of vacuum, $r$, $d$ and $w$ respectively denote the radius of the internal ring, the distance between the internal and external rings, and the width of the rings (see Fig.~\ref{cpwsrr}), while the arguments of the elliptic integrals $K$ and $K'$ are now given by:
\begin{eqnarray}
\kappa_3=\frac{d}{d+2w}, \qquad
%\label{k0}\\
\kappa_4=\frac{\sinh\left(\frac{\pi d}{4h_1}\right)}{\sinh\left[\frac{\pi (d+2w)}{4h_1}\right]}.
\label{k34}
%\\k_2&=&\frac{\sinh(\frac{\pi W}{4h_2})}{\sinh(\frac{\pi (W+2G)}{4h_2})}\label{k2}
\end{eqnarray}

\section{The perturbation scheme}

As mentioned in Sec.~II.B, the quasi-continuum approximation is a variant of the method of multiple scales \cite{Jeffrey}. We thus introduce, at first,
a set of new independent temporal variables, $t_n = \epsilon^n t$ ($n=0,1,2,\cdots$), and acordingly expand the derivative operator
$\partial_t$ as $\partial_t = \partial_{t_0} + \epsilon \partial_{t_1} + \cdots$.
Next, we seek solutions of Eq.~(\ref{eq:model}) in the form:
\begin{equation}
V_n =\epsilon u_{1n}(t_n)e^{i\theta_n} +\epsilon^{2} u_{2n}(t_n)e^{i2\theta_n}+\cdots +{\rm c.c.},
\label{eq:ansatz01}
\end{equation}
where $\theta_n = \omega t_0 - k n$. Then, we substitute Eq.~(\ref{eq:ansatz01}) into Eq.~(\ref{eq:model}) and employ a continuum approximation for the unknown envelope functions $u_n$, namely $u_n \rightarrow u(x)$, where $x=n\alpha$ and $\alpha$ being the lattice spacing (the latter parameter does not appear in the results below, as one may readily rescale $x$ as $x/\alpha$). Furthermore, similarly to the introduction of the temporal variables, we introduce the set of the spatial variables $x_n = \epsilon^n x$ (and, thus, $\partial_x = \partial_{x_0} + \epsilon \partial_{x_1} + \cdots$). To this end, equating coefficients of like powers of $\epsilon$, we obtain the following (first three) perturbation equations:
\begin{eqnarray}
&O(\epsilon )&: \qquad \hat{L}_0 u_1=0, \\
\label{eq:term1}
&O(\epsilon ^{2})&: \qquad \hat{L}_0u_2+ \hat{L}_1u_1= \hat{N}_0 u_1^2,  \\
\label{eq:term2}
&O(\epsilon ^{3})&: \qquad \hat{L}_1u_2+ \hat{L}_2u_2= \hat{N}_0[u_1u_2],
\label{eq:term3}
\end{eqnarray}
where the operators are given by
\begin{eqnarray}
\hat{L}_0 &=&\frac{\partial^4}{\partial t_0^4}+\left(1+\gamma^2+4\beta^2\mu^2\sin^{2}\frac{k}{2}\right)\frac{\partial^2}{\partial t_0^2} \nonumber \\
&+&4\beta^2\mu^2\sin^{2}\frac{k}{2}+\gamma^2,  \\
\label{eq:LO}
\hat{L}_1&=&\frac{\partial^4}{\partial t_0^3 \partial t_1}+2\left(1+\gamma^2+4\beta^2\mu^2\sin^{2}\frac{k}{2}\right)\frac{\partial^2}{\partial t_0 \partial t_1}\nonumber\\
&+&2i\beta^{2}\sin{k}\frac{\partial^3}{\partial t_0^3 \partial x_1}, \\
\label{eq:L1}
\hat{L}_2&=&6\frac{\partial^4}{\partial t_0^3 \partial t_1}+2\left(1+\gamma^2+4\beta^2\mu^2\sin^{2}\frac{k}{2}\right)\frac{\partial^2}{\partial t_1}\nonumber\\
&+&4i\beta^{2}\sin{k}\frac{\partial^3}{\partial t_0 \partial t_1 \partial x_1}+2i\beta^2\mu^2\sin{k}\frac{\partial}{\partial x_1},
\\ \label{eq:L2}
\hat{N}_0&=&\frac{\partial^4}{\partial t_0^4}+\gamma \frac{\partial^2}{\partial t_0^2}.
\label{eq:N0}
\end{eqnarray}

It is clear that the first-order pertubation solution to Eq.~(\ref{eq:term1}) reads:
\begin{equation}
u_1=V_1(x_1,x_2,\cdots,t_1,t_2,\cdots)\exp(i\theta) +{\rm c.c.},
\label{eq:u1}
\end{equation}
where $V_1$ is an unknown complex function, $\theta = \omega t_0 - k x_0$, while $\omega$ and $k$
satisfy the dispersion relation of Eq.~(\ref{eq:dispqc}). Next, substituting Eq.~(\ref{eq:u1}) into
Eq.~(\ref{eq:term2}), we obtain the following results: first, the non-secularity condition:
\begin{equation}
\frac{\partial V_1}{\partial t_1}-\left[\frac{\beta^2\sin k(\omega^{2}-\mu^{2})}{2\omega^{3}-(1+\gamma^{2}+4\beta^2\sin^{2}\frac{k}{2})\omega}\right]\frac{\partial V_1}{\partial x_1}=0,
\label{eq:secular}
\end{equation}
suggests that $V_1=V_1(X,x_2,\cdots, t_2,\cdots)$, where $X=x_1-v_g t_1$ and $v_g$ is the group velocity of Eq.~(\ref{eq:gvelqc}), now consistently determined; second, we obtain a uniformly valid solution for the second-order perturbation equation, in the form:
\begin{equation}
u_2= \frac{4\omega^{2}(\gamma^{2}-4\omega^{2})}{{\mathcal{G}(\omega,k)}} \exp(i2\theta)+{\rm c.c.},
\label{eq:u2}
\end{equation}
where $\mathcal{G}(\omega,k)$ is given by Eq.~(\ref{eq:Gparam1}) and the dependence of $u_2$ on higher-order scales has been omitted. To this end, substituting Eqs.~(\ref{eq:u1})-(\ref{eq:u2}) into Eq.~(\ref{eq:term2}), and using the variables $X=x_1-v_g t_1 \equiv \epsilon(n-v_g t)$ and $T=t_2 \equiv \epsilon^2 t$, we derive from the non-secularity condition at
$O(\epsilon ^{3})$ the NLS Eq.~(\ref{eq:NLS}).

%\appendix
\section{The continuum approximation}

In the continuum limit (for $k \rightarrow 0$), the dispersion relation [cf. Eq.~(\ref{eq:dispqc})]
is reduced to the form,
\begin{equation}
\omega^{4}-(1+\gamma^2+\beta^2k^2)\omega^2+\beta^2\mu^2k^2+\gamma^2=0.
\label{eq:dispc}
\end{equation}
The group velocity, $v_g = \partial \omega/\partial k$, is now given by
\begin{equation}
v_{g}=\frac{\beta^2k(\omega^{2}-\mu^{2})}{2\omega^{3}-(1+\gamma^{2}+\beta^2k^{2})\omega}.
\label{eq:gvelc}
\end{equation}
Finally, the expressions for the dispersion and nonlinearity coefficients $P$ and $Q$ in the continuum approximation read:
\begin{eqnarray}
P&=&
\frac{(1+\gamma^{2}+\beta^2k^{2}-6\omega^{2})v_{g}^{2}}{-4\omega^{3}
+2(1+\gamma^{2}+\beta^2k^{2})\omega}\nonumber\\
&+&
\frac{4\beta^2\omega k v_{g} +\beta^2(\omega^{2}-\mu^{2})}{-4\omega^{3}+2(1+\gamma^{2}+\beta^2k^{2})\omega},
\label{eq:dispcoefc}
\end{eqnarray}
and
\begin{equation}
Q=\frac{4\omega^{4}(\gamma^{2}-\omega^{2})(4\omega^{2}-\gamma^{2})}{[-2\omega^{3}
+(1+\gamma^{2}+\beta^2k^{2})\omega]\mathcal{G}},
\label{eq:nlcoefc}
\end{equation}
where the function $\mathcal{G}$ is now given by:
\begin{equation}
\mathcal{G}=16\omega^{4}-4(1+\gamma^{2})\omega^{2}+\gamma^{2}-4\beta^2(4\omega^{2}-\mu^{2})k^{2}.
	\label{eq:Gparam2}
\end{equation}

\end{document}